   \documentclass{emulateapj}
\usepackage{natbib}

 \shortauthors{Jourdain et al.}
\shorttitle{High Energy Emission of V404 Cygni during the 2015 outburst finale  with \textit{INTEGRAL}/SPI}

\begin{document}

\title{A challenging view of the 2015 summer V404 Cyg outburst at high energy with \textit{INTEGRAL}\footnotemark[1]/SPI: The finale.
}
\footnotetext[1]{Based on observations with \textit{INTEGRAL}, an ESA project with instruments and science data centre funded by ESA member states
 (especially the PI countries: Denmark, France, Germany, Italy, Spain, and Switzerland), Czech Republic and Poland with participation of
 Russia and USA.}
 
\author{Elisabeth~Jourdain\altaffilmark{1}, Jean-Pierre~Roques\altaffilmark{1}, James Rodi\altaffilmark{1}\\
}


\affil{\textsuperscript{1}Universit\'e Toulouse; UPS-OMP; CNRS; IRAP; 9 Av. Roche, BP 44346, F-31028 Toulouse, France\\
}

\begin{abstract}

During its strong outburst of 2015 June/July, the X-ray transient V404 Cygni (= GS2023+338) was observed up to a level of 50 Crab in the hard X-ray domain. We focus here on a particularly intense episode   
preceeding a definitive  decline of the source activity.
We benefit from large signal-to-noise ratios to investigate the source spectral variability on a timescale of 5 minutes. A Hardness-Intensity study of three broad bands reveals clearly different behaviors at low and high energy (below and above  $\sim$ 100 keV). In particular, 
on two occasions, the source intensity varies by a factor of 3-4 in amplitude while keeping the same spectral
 shape. On the other hand, at the end of the major flare, the emission presents a clear anticorrelation between flux and hardness.
These  behaviors strongly suggest  the presence of two spectral components related to emission processes varying in a largely independent way. The
first component (E $<$ 100-150 keV) is classically identified with a Comptonizing thermal electron population and requires either an unusual seed photon population
or a specific geometry with strong absorbing/reflecting material.
The second component is modeled by a cutoff power-law, which could correspond 
 to a second hotter Comptonizing population or another mechanism (synchrotron, non-thermal Comptonization...).
In the framework of such a model,  Hardness-Intensity and Flux-Flux diagrams 
clearly demonstrate that the source evolution follows a well organized underlying  scheme. They reveal unique information about
 the Hard X-ray emission processes and connections between them.

\end{abstract}      


\maketitle

\section{Introduction}

X-ray binaries (XRBs) emit a large fraction of their luminosity in the high energy domain. Observing the
source behavior above a few keV is thus a prime importance. Up to a few keV, most of the emission comes from 
the accretion disk but above $\sim$ 10 keV, the emission is thought to  come from a hot corona, where 
 electrons, with a temperature of 20-60 keV, Comptonize the disk photons. 
Transient XRBs containing a black hole represent a subsample of XRBs which can reach, for a short time interval, high flux levels, allowing deep studies of the emission mechanisms.
The best example has probably happened recently, with the outburst of V404 Cygni (=GS2023+338) in 2015 June/July (Barthelmy et al. 2015 , Younes, 2015). The source luminosity  presents a huge variability for $\sim$ 12 days, 
with a series of flares of various intensities. The last one is the most impressive, with a flux above 20 Crab in the hard-X ray domain for 
 more than 8 hours  (around MJD 57199.7, $\sim$11 days after the beginning of the outburst).
Thanks to exceptional signal-to-noise ratios, the data bring us invaluable information about the spectral evolution on  timescales
 as short as 5 minutes.  We provide in this paper a detailed study of the evolution of the source emission, from 20 keV to $\sim$ 600 keV during the INTEGRAL revolution
 number 1557 ($\sim $ 2 days duration), where the source brightness reaches its maximum. We first present briefly the instrument and
 the observations. The source behavior is then studied through Hardness-Intensity Diagrams (HID) built on a 5 minute timescale. The decomposition of the high energy emission into two components leads to an unified description of 
both the various spectral shapes and the global comportement of V404 Cyg above 20 keV, during this extraordinary event.

\section{Instrument, observations and data analysis}\label{instru}
\subsection{Brief description of the SPI spectrometer}
The INTEGRAL mission was launched in October 2002 from Ba\"{\i}konour, Kazakhstan. The SPI instrument is a spectrometer based on a Ge camera associated with a coded mask located 
1.8 m above the detector plane. It provides  images of the sky in the energy range from 20 keV to 8 MeV with a modest spatial resolution (2.2$^\circ$) and a large field of view 
(30$^\circ$). The detector plane is made of 19 crystals of high purity Germanium operating at 80 K in order to offer an energy resolution from 2 to 8 keV in the whole energy domain. BGO blocks surround  the camera to protect it from ambiant particles. They also measure the background photons and remove them from the signal. 
The INTEGRAL mission is described in more detail in Winkler et al. (2003), the SPI instrument design in Vedrenne et al. (2003) and the SPI performance in Roques et al. (2003).
\\
\subsection{Observations and Data analysis}
Since 2015 February, the INTEGRAL observation schedule consists of 2.5 day revolutions (with instruments off for 0.5 day below radiation belts) split into $\sim$ 50 exposures or science window (scw) of $\sim$ 3000 s. The pointing direction is shifted by 2.2 $^{\circ}$ from one scw to the next one, to follow a regular dithering pattern (5X5 rectangular or 7-Hexagonal).\\
In this paper, we analyzed the data recorded during the revolution 1557, starting on 2015 June, 26 (01:44) and including the major flare of the activity period. Due to the huge source flux, some telemetry packets are lost. It concerns only the few hours around the maximum and represents less than 10 \% of the data.
These telemetry gaps are taken into account in the determination of the useful duration, supplementing the usual electronic  deadtime and do not represent an issue for the spectral analyses.
To get the source information, the detector count rates are adjusted with
a sky model through a $\chi2$ minimization method by a model fitting procedure (see Jourdain \& Roques, 2009) for a detailed description).
In addition to V404 Cyg, the sky model includes Cyg X-1 and Cyg X-3,  with  their flux normalization allowed to vary respectively on  $\sim$ 10 hours and one revolution timescales.  The source fluxes are extracted for each time bin and each energy bin.
An event selection  criterion based on a 'high energy' flag has been used for the energies between 400 keV and 2.2 MeV.
Indeed, a spurious noise is known to  produce artefacts above $\sim$ 700 keV. Furthermore, due to pile-up between the huge source photon flux and the pulse tail of  high energy saturating events, a tiny fraction of the source photons are recorded at a wrong (higher) energy.
Using the  'high energy' flag (which has a threshold of  $\sim$ 300 keV) as a confirmation for E $>$ 400 keV photons
suppresses these wrong events.
A correction factor of 1/0.85 is applied to the counts extracted 
in this energy band to take into account the efficiency loss induced by the 'high energy' flag selection (Roques \& Jourdain, 2016). 
Revolutions 1546 to 1549 are considered as 'empty field' and used to estimate the background pattern (relative detector normalisation map). 
During the model fitting procedure, the amplitude of the background has been allowed to vary on the scw timescale (i.e. $\sim$ 1 hour). 

For the following study, broad band light curves  have been built by extracting count rates in  three 3 energy bands from 20 keV to 300 keV 
(F1=$F_{[20-50]}$; F2=$F_{[50-100]}$; F3=$F_{[100-300]}$), on a 300 s timescale (Fig. \ref{fig:LC}).
Count spectra are extracted in 43 logarithmically spaced energy bins from 20 keV to 1 MeV,
and deconvolved with the associated response through the {\sc xspec} (v12.8.2) tools. 
Below 25 keV, uncertainties on the instrument responses may lead to significant deviations in the spectral residuals for any of the tested standard spectral models, and we ignore  them in the presented results.

 \section{Light curves and Hardness - Intensity Diagrams}\label{HID}
 
\begin{figure}[l]
 \includegraphics[width=40mm,height=10cm, angle=90,]{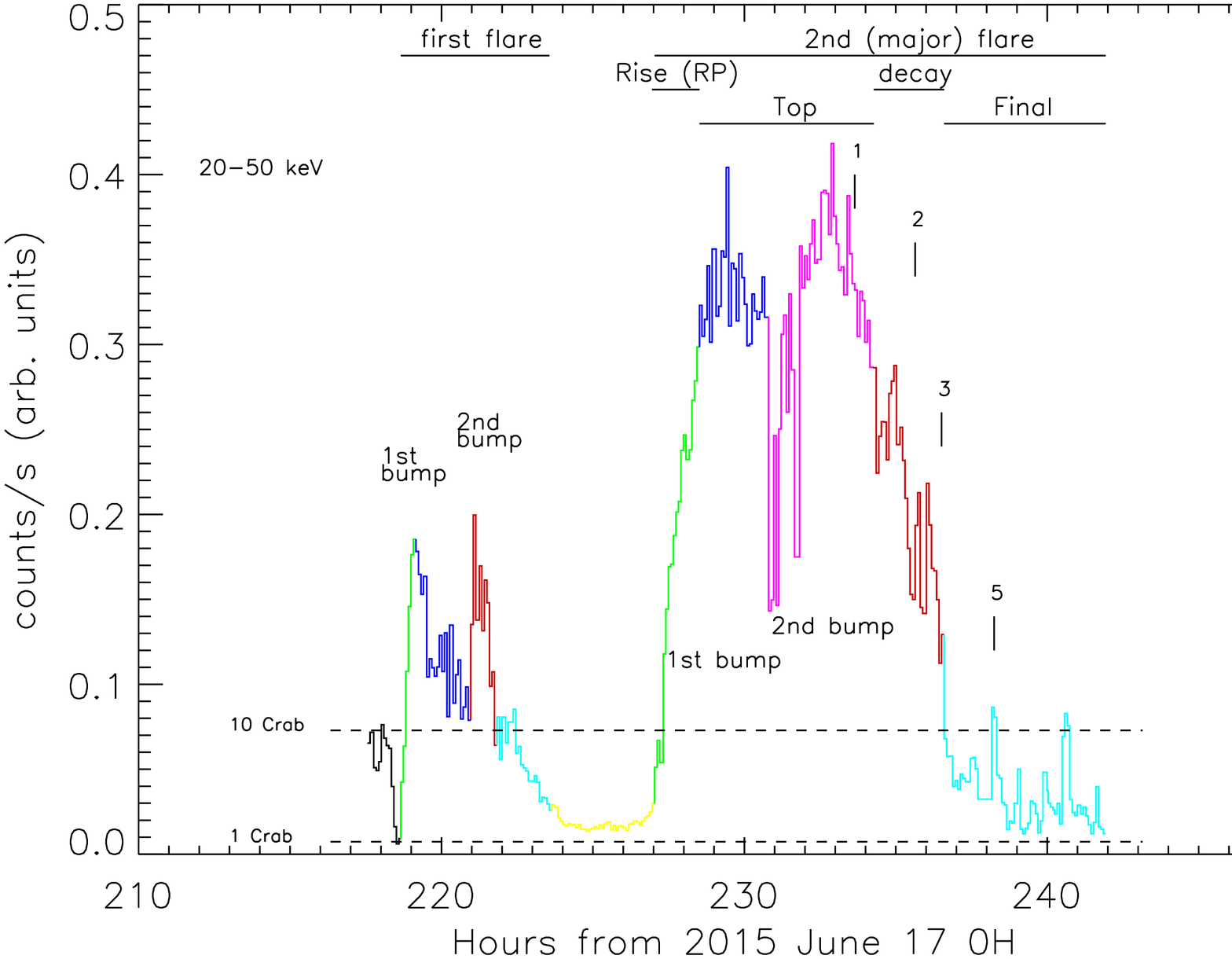}\\
  \includegraphics[width=40mm,height=10cm, angle=90,]{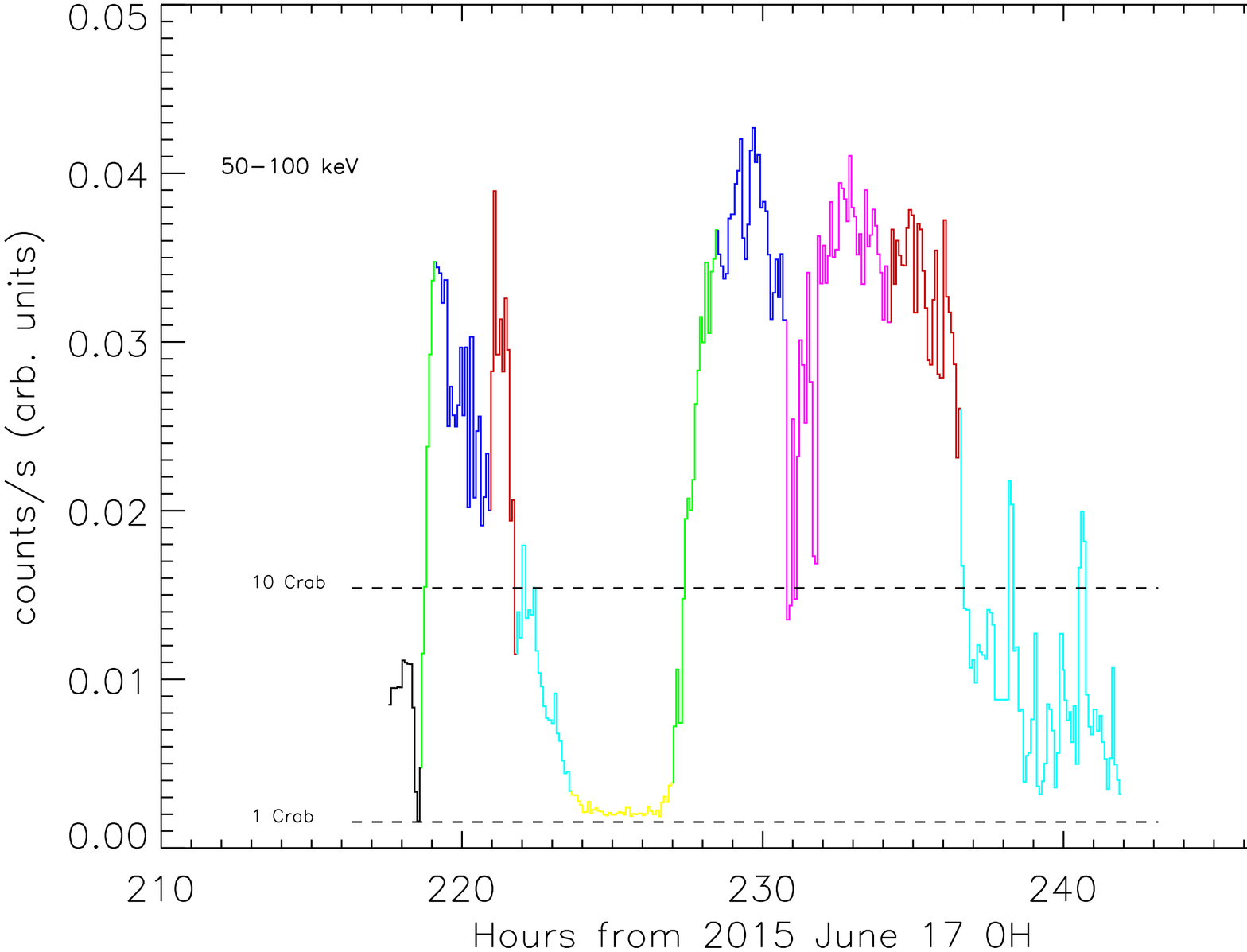}\\
  \includegraphics[width=40mm,height=10cm, angle=90,]{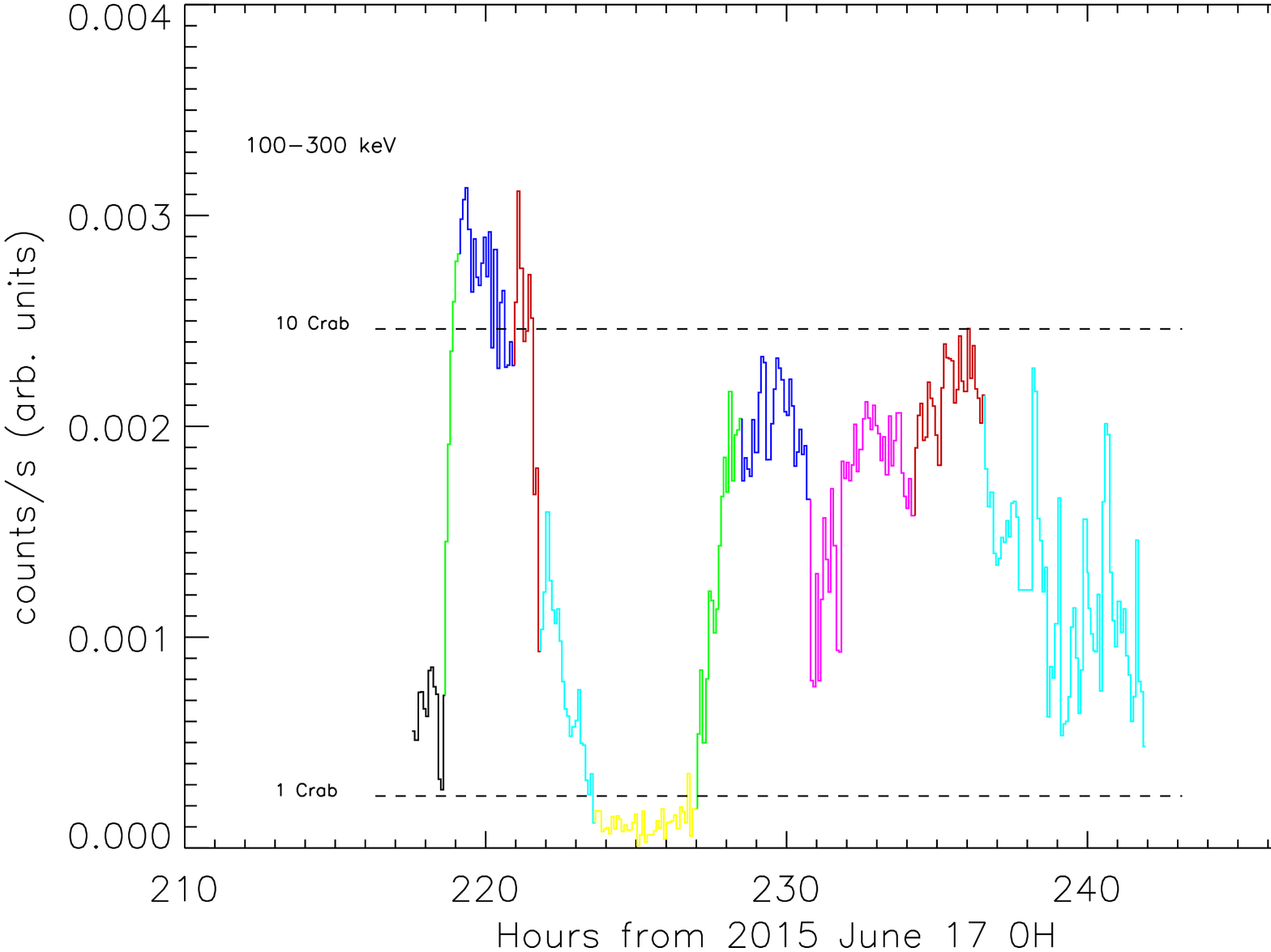}\\
 \caption{V404 Cyg light curve along Rev. 1557, for E=[20-50 keV], [50-100 keV] and [100-300 keV]
from top to bottom. Time bins=300 s}
\label{fig:LC}
\end{figure}

Fig. \ref{fig:LC} illustrates the time evolution of the source in the three considered broad bands. The source flux displays a high variability, even above 100 keV. It exceeds  50 Crab in the
 20-50 keV energy range and  10 Crab in the 100-300 keV one. The presence of impressive "jumps" on the 5 minute timescale suggests significant variability on  shorter timescales.
 For instance, at the beginning of the second bump of the second flare (pink light curve part), a flux drop corresponding to about 25 Crab (20-50 keV band) is visible, followed, $\sim$ 1 hour later, by an increase of similar amplitude, both in less than 300 seconds.

\begin{figure}
 \includegraphics[width=8cm,height=8cm, angle=90]{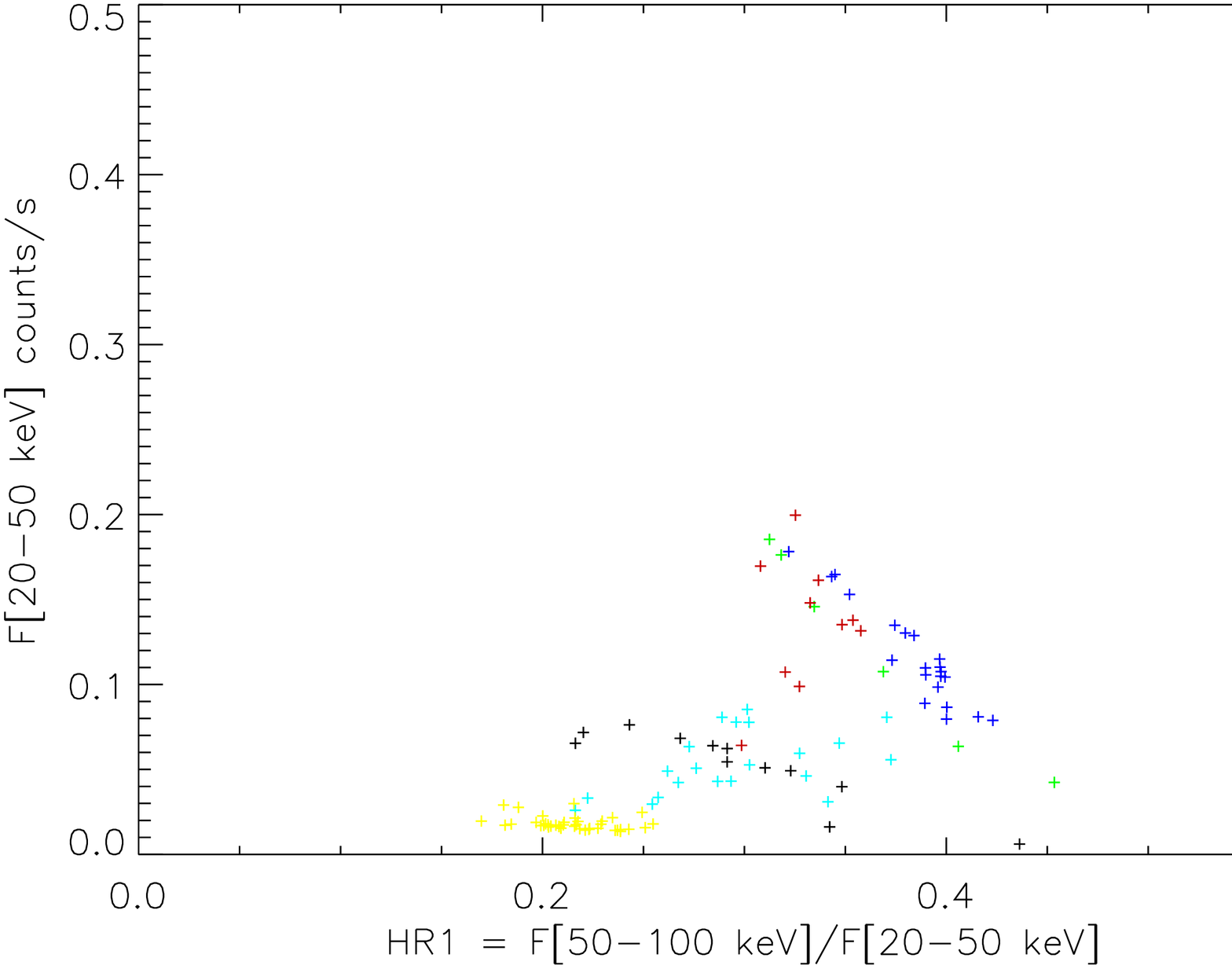} 
 \includegraphics[width=8cm,height=8cm, angle=90]{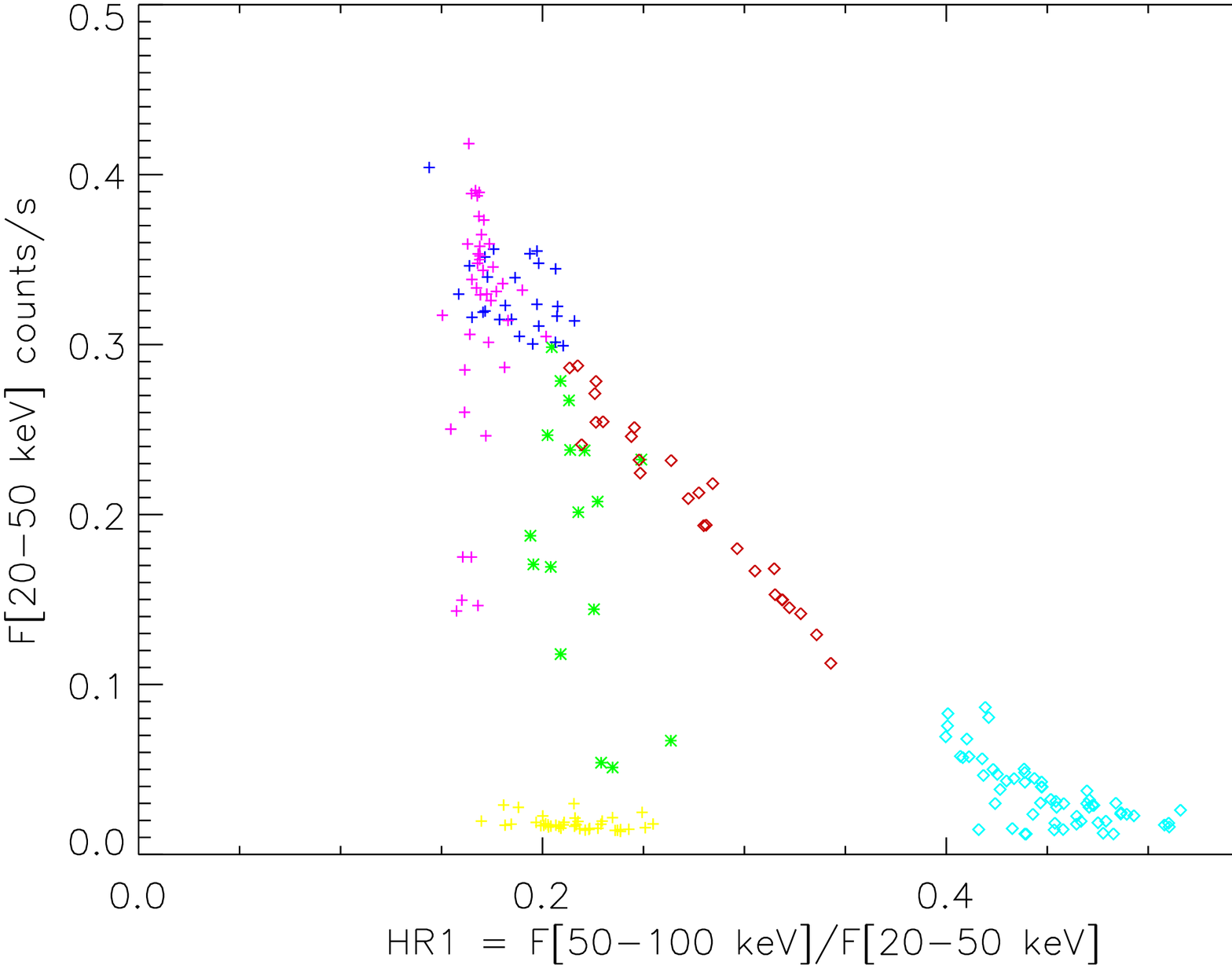}
 \caption{The Hardness-Intensity Diagram (HID) at low energy (F1 vs H1=F2/F1 ) for the first  flare [left]
and the second flare [right] of revolution 1557. }
\label{fig:HID1}
\end{figure}

The source evolution  differs strongly from one energy band the other.
While two broad structures (1st flare, 2nd flare), both showing a double bump profile,  appear whatever the energy band,  their relative importances evolve drastically.
Twice as faint as the second flare at low energy, the first flare remains above 10 Crab for E $>$
 100 keV, where it dominates the light curve profile. This points to the complex behavior of the source emission, which changes from one flare to the next one.\\
As a first step in the spectral evolution study, we have built two hardness ratios: HR1=F2/F1 and HR2=F3/F2 and studied them versus the source intensity
(through Hardness-Intensity Diagrams, hereafter HID).  Note that the color code in the HID is the same as used in Fig.\ref{fig:LC}. 
Fig.\ref{fig:HID1}  displays the HIDs corresponding to the low energy part (HR1=f(F1)). 
 For clarity, we plot separately the data for the first (left panel) and second (right panel) flare.\\ 
The HID evolution presents the most striking pattern during the major flare: Starting from a low flux level,
the source flux increases by a factor 3-4, at almost constant (low) hardness (yellow stars). After the rise phase, the first plateau ('first bump' in Fig.\ref{fig:LC})) appears without any significant evolution (blue points). Then, a large drop in flux  corresponds to the start point of a new episode of high flux variability at constant hardness (violet points), covering  what appears as a second bump in the light curve. The emission is softer than the rise phase, due to the 20-50 keV flux which reaches its maximum at this moment.
After these two episodes where the source emission varies mainly in amplitude but not in shape, a spectral evolution is triggered, 
and we observe, in particular at low energy, a decay phase exhibiting a strong anti-correlation between flux and hardness.
This tight relation is preserved from the maximum up to a low flux level phase, corresponding to the end of the most  active phase of the source.
Note that the continuously hardening decay spans a flux amplitude similar to that of the spectrally stable rise phase, but is much more structurated.\\
After MJD 57199.85 (T0+236.85 H in Fig.\ref{fig:LC}), the source remains in a 'hard state' (proper to this source), with a  moderate  flux level (still a few Crabs) but still large flux variability.\\
Comparing the HIDs of the first and second flares shows that, during the first flare, the source remains localized
 in the hardest part of the triangle shaped second flare HID, and exhibits  at times a similar anti-correlation behavior.

\begin{figure}
 \includegraphics[width=8cm,height=8cm, angle=90]{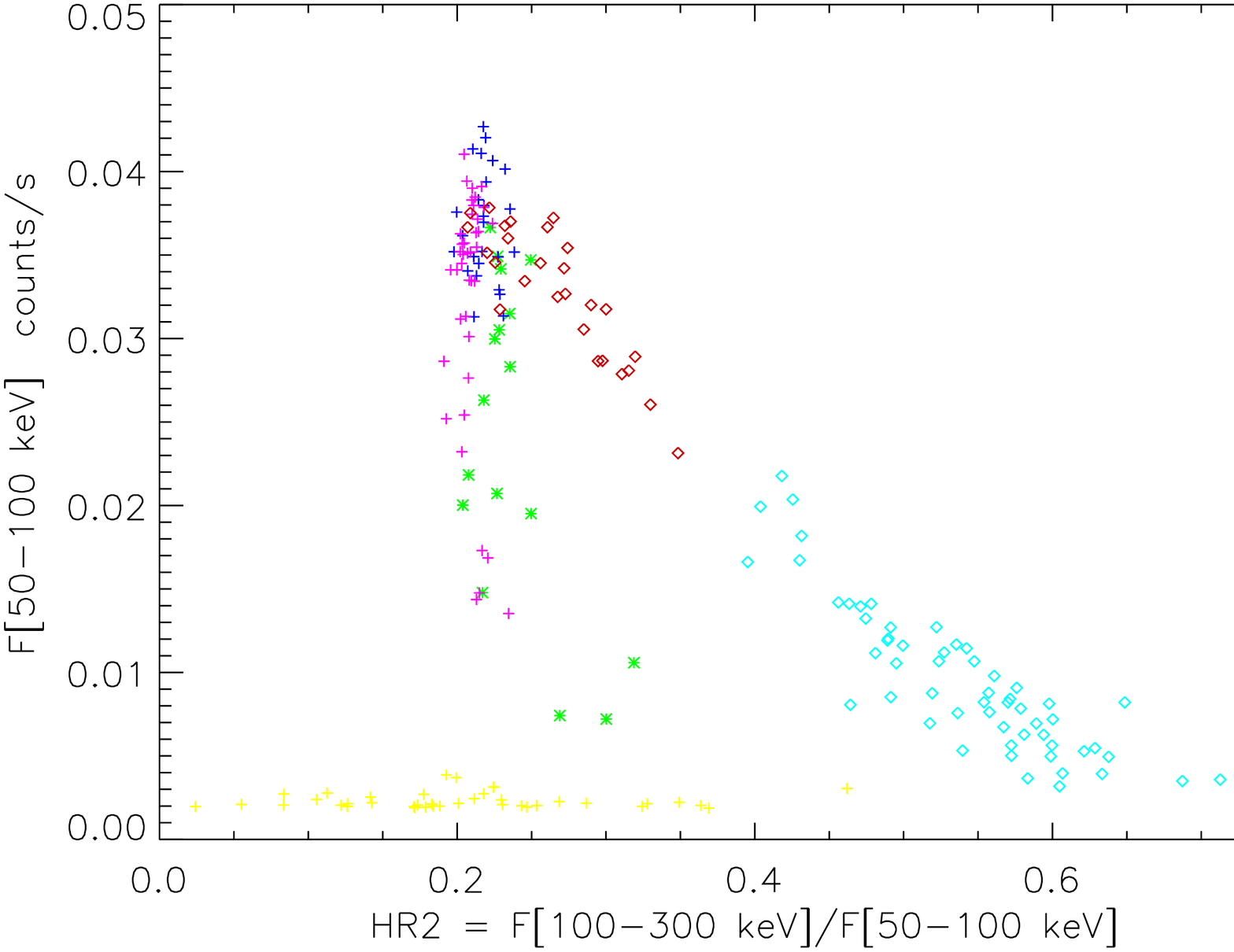} 
  \includegraphics[width=8cm,height=8cm, angle=90]{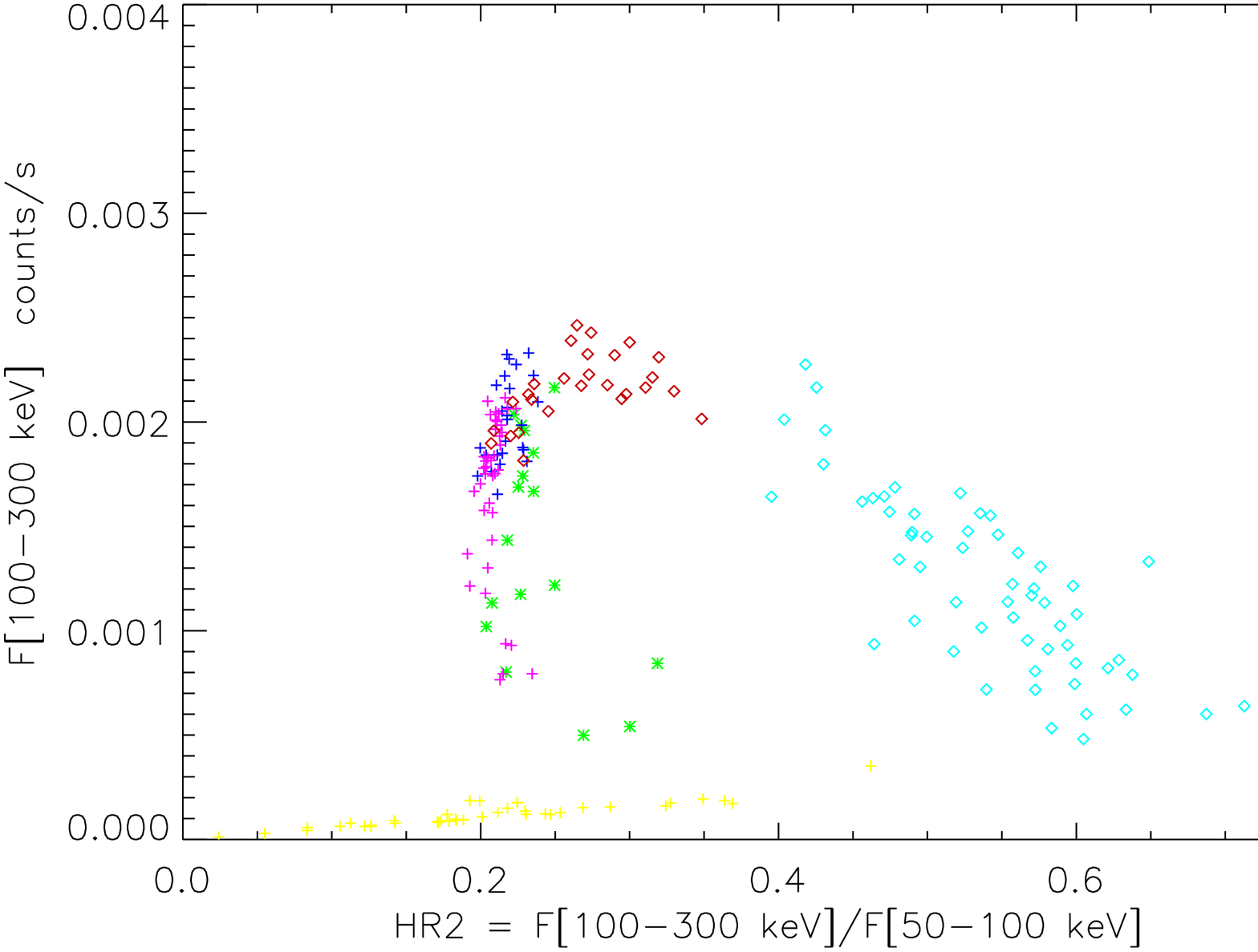}
\caption{HID at higher energy, for the second flare only. 
F2 vs H2=F3/F2 [left] and F3 vs H2 [right].}
\label{fig:HID2}
\end{figure}

To investigate the source emission at higher energy,  we used the hardness ratio HR2= F3/F2 (above and below 100
 keV) and display in Fig. \ref{fig:HID2} the source hardness evolution as a function of two different fluxes, during the main flare.
The left corresponds to HR2= F3/F2 versus F2 and right one to HR2 versus F3. 
When plotting H2=f(F2), the HID pattern is similar to that observed at low energy (H1 versus F1). The only noticeable difference is that
the two periods with variable flux and constant hardnesses  (rise phase and second bump) follow the same track (same H2 hardness).  
More striking is the pattern observed in the second panel, where H2 is plotted as a function of the high energy flux (100-300 keV).
The triangle shaped  evolution drawn at low energy is still suggested. However, the flux evolution during the flare is confined in a much smaller
amplitude range, as if the source emission was saturating. In this energy range, the 
extreme flux values during the top of the flare are  comparable to those observed in the last part
of the observation.
Consequently,  the flux decay observed at low energy does not exist above 100 keV.
Instead, the source keeps a roughly constant flux while going from the low to high hardness values, implying a decrease of the low energy emission.
This reflects the flat profile observed in the 100-300 keV light curve (Fig.\ref{fig:LC}). 
Finally, we notice that an anticorrelated behavior is observed during the postflare period, whatever the considered energy band.

All the particularities observed above strongly suggest that at least two different mechanisms evolving independently 
are required to interpret the hard X-ray  emission of V404 Cyg (E $>$  20 keV) during the outburst. 
  This fits perfectly with previous results reported for V404 Cyg itself (Rodriguez et al. 2015, 
 Natalucci et al. 2015 and Roques et al. 2015) and other X-ray binary sources like Cyg X-1.
Such analysis allows quantifying the energy releases underlying the source emission. Its major advantage is
that it is model-independent and reveals the evolution pattern on timescales usually unreachable at these energies.

However, we can try to better understand the respective role of these two components, through 
a spectral analysis, as presented in the next section.

 \section{spectral study}\label{spectral}

To explore more precisely the spectral evolution of the source, the observations can be compared to  physical models.
Based on the conclusion of the previous section,  the emission is described with two components and the simplest usual scenario involves
 a Comptonization,  which is the classical mechanism to explain the X-ray binary emission between 20 and $\sim$ 100 keV, 
and a cutoff power-law to account for the high energy part (CompTT +cutoffpl in xspec language, CompTT from Titarchuk, 1994). Unfortunately,  this very basic scheme recovers a large number of possible 
pictures, which will give similar spectral shapes. The following discussion is thus based on some hypotheses we state below and 
 the values of the deduced parameters are to be considered as more indicative than absolute.
First, to limit the unavoidable degeneracy between the free parameters, we  fix  the cutoff
 power-law index to 1.6   and the cutoff energy to 150 keV, when not constrained by the data.
Generally, we aim to find the best compromise between acceptable $\chi2$ values and a minimal number of 
degrees of freedom, by imposing (when possible) common parameter values among the spectra.
Moreover, we have observed (see also previous works by Roques et al. 2015, Natalucci et
al. 2015 , and Jenke et al. 2016), that the low energy spectral shape can not be adequately fit with usual parameter values:  
 it requires huge values of density column/ reflection factor or a "hot" seed photon temperature 
(5-7 keV).  
None of these solutions are satisfying and we opted  (as in previous works) for a model with a hot seed photon population, considering that
it could mimic the presence of another physical process  (seed photons from synchrotron emission, ...)
or a peculiar geometry (absorbing or reflecting medium, etc.; see Discussion Sect.\ref{discussion}).
 
\begin{figure}
\includegraphics[width=8cm,height=8cm,angle=0]{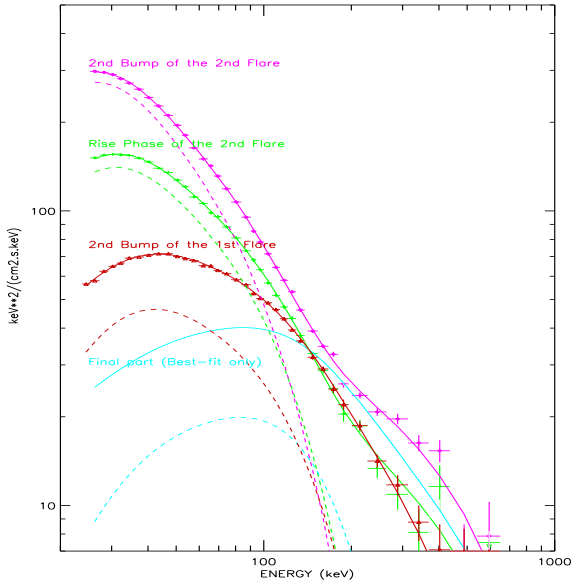} 
\caption{ Comparison of Hard and Soft "states" of V404 Cyg as observed during Rev 1557. The three brightest spectra illustrate the source evolution between the 2nd bump of the first flare, and rise phase and 2nd bump  of the second flare. Spectrum averaged over the last (hardest) part of the 2nd flare is shown for comparison.  
Dashed lines correspond to the respective Compton components and illustrate their role in the source evolution.} 
\label{fig:SoftHard}
\end{figure}

\begin{figure}
\includegraphics[width=8cm,height=8cm,angle=0]{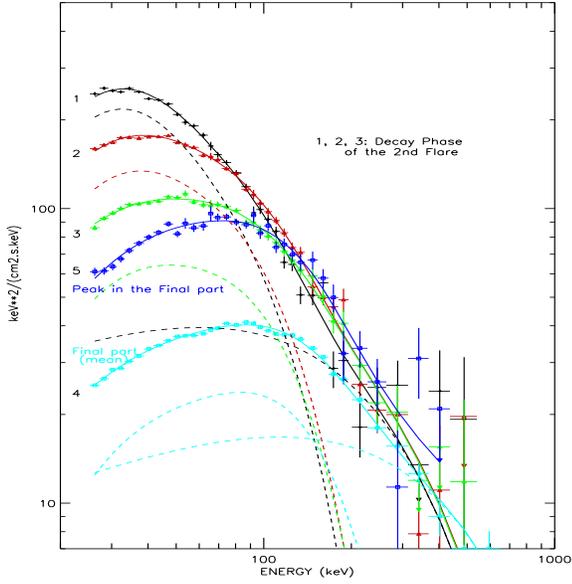}\\
\caption{Spectral evolution of V404 Cyg during the decay of the 2nd flare (1-3) and the final part (4-5).  
Dashed lines correspond to the respective Compton components. Cutoff power-law has a fixed cutoff energy for the three
 decaying spectra (shown only for the highest one) and free (but common to both) for the two final spectra 
(see text and Tab.\ref{tab:fits}).} 
\label{fig:decay}
\end{figure}

Based on this two component model, we are able to follow the source parameter evolution along the outburst. 
As a complete description of the 280 spectra  corresponding to the 300 second intervals used in the light curves and HIDs presented above
is out of the scope of this paper, we have chosen a few spectra representative of the different 'states' identified from the HIDs to illustrate 
the respective role of both components.
 Results are displayed in Fig.\ref{fig:SoftHard} and  \ref{fig:decay} and Table \ref{tab:fits}.
 
\begin{itemize}
\item{First Flare:}\\
 Even though the source spectral shape cannot be considered as constant, the emission remains 
 in a limited area in the HID. We have built the spectra corresponding to the four parts identified 
in Fig.\ref{fig:LC} along this first flare. When fit simultaneously, they  can be  described with the 
same best fit parameters (kT $\sim$ 25 keV, Ecut = 150 keV), except the optical depth which is required to evolve between 1.3 and 2.1 
(1.6, 2.1, 1.6, 1.3 respectively from beginning to end).  In Fig.\ref{fig:SoftHard}, this episode is represented
by the spectrum averaged over the second bump (8 ks), which presents a similar flux level that the spectrum averaged over the second flare rise phase above $\sim$ 150 keV.

\item{Rise phase and top of the second flare:}\\
 Considering the periods which present a constant hardness (and thus plausibly a constant spectral shape), we have summed up all the corresponding spectra to get a high signal-to-noise ratio spectrum. \\
We thus get mean spectra for the rise phase (RP) 
and second bump of the second flare. Between these two periods, the source emission softens at low energy (HR1), while keeping the same HR2.
In terms of spectral parameters, the Comptonizing population becomes thiner ($\tau$=1.3 to 1.0). Interestingly, the electron temperature is not required to change but the seed
photon temperature is. 
In addition, we have observed  that during the first bump (which joins the rise phase and the second bump), as expected from the HIDs, the emission behaves precisely as a passage
 from one to the other, with parameter values in between those mentioned above.\\
In Fig.\ref{fig:SoftHard}, we compare these two spectra with the emission of the first flare. This points
out how the source evolution is almost exclusively driven by the Compton component. 

\item{Decay phase:}\\
Fig.\ref{fig:decay} displays the results for a sample of spectra taken along the decay (labels 1-3). 
Based on fitting the three spectra together with the seed photon temperature linked, a common value of 6.35 keV
was found. Leaving the individual values free does not improve the $\chi2$ values. From Table \ref{tab:fits}, we conclude that 
here too, the continous hardening of the spectral emission below 100 keV is mainly due to an increase of the optical depth from 1.2 to 2.1. 
Due to the small duration of individual spectra (300 s), the cut-off energy cannot be constrained and the second component shape is  thus fixed ($\alpha$ = 1.6 and $E_{cut}$= 150 keV). Its normalization slightly
decreases with the global source flux, but its relative contribution increases since the Compton component drops by a factor of $\sim 3$.

\item{Final phase:}\\
During the last part of the observation, the spectral emission remains in a hard state which extends the evolution observed along the decay.  
We see that the source still exhibits significant variability both in intensity and hardness, probably driven by the low energy component. In Fig.\ref{fig:decay}, the averaged spectrum (label 4) is displayed 
together with a spectrum corresponding to a small peak observed at T0+238~H (label 5). The later reveals   additional emission at low energy, due to an increase of the Compton component.\\
To compare the emission observed during this time interval to the hard emission
observed during the first peak, we have plotted on Fig.\ref{fig:SoftHard} the best-fit of the mean spectrum ($N^0$4). It is clear that 
 the final phase components are both significanlty harder. In terms of spectral parameters, this translates 
into higher kT, $\tau$  and $E_{cut}$  values (kT $\sim$ 30 keV, $\tau$ $\sim$ 2, $E_{cut}$ $\sim$ 200 keV).
The  Compton contribution weakens and, at low flux levels, it could be that the cutoff power-law component is 
responsible for most of the emission over the entire energy domain. 
\end{itemize}

\section{Discussion}\label{discussion}
V404 Cyg is a transient Low-Mass X-ray binary which both is the brightest system in quiescence (see, for instance, Reynolds et al. 2014) and reaches the highest intensity levels during active phases.
It is  located at 2.4 kpc (Miller-Jones et al. 2008), a distance comparable to that of the brighest 
permanent hard X-rays sources, Cyg X-1 and the Crab Pulsar/Nebula, both emitting at a  $\sim$ 1 Crab level in the hard X-ray domain,
V404 Cyg appears thus 50 times brighter than these famous persistent sources. On  the other hand, if located 
in the Galactic Center, V404 Cyg would have reached about 3 Crab at its maximum, and presented a significant level of activity
(above $\sim$ 100 mCrab up to 100 keV) during more than 12 days. The flaring episode reported here would have been detected at
 $\sim$  1 Crab for about 10 hours. This outburst from V404 Cyg is thus clearly an unprecedented event. 
Such high fluxes are more common in Be systems,  when the central neutron star passes
through the companion star's envelop. In this case, the huge luminosity is thermally emitted in the soft X-ray band (below 10-20 keV).
In the case of the V404 Cyg system, most of the luminosity is emitted above 20 keV, implying a different nature of the mechanisms at work.
It suggests that Comptonization (or eventually synchrotron) plays a major role in the energy release, which reaches here spectacular levels. By analogy with other binary systems, the source has to be classified in a "low Hard State", as observed for Cyg X-1 for instance. However, the V404 Cyg emission is significantly softer than in Cyg X-1, particularly in the hard tail emission (cutoff around 200 keV instead of 500-700 keV). Nevertheless, 
we have demonstrated that V404 Cyg presents a large variability in terms of emission hardness and we have 
identified 'Hard' and 'Soft' states, which did not refer to the usual terminology. 

The Hardness-Intensity correlation is usually built in a lower energy domain, and the cross-evolution of
the disk emission around/below a few keV  and a hard X-rays emission (attributed to a corona) from a few up to $\sim$ 20 keV is studied.
In this energy domain, HIDs often present a typical Q-shape. Our study is dedicated to the hard X-ray domain and addresses quite different physics.
It is thus not surprising to get unusual HIDs.\\
The two main tracks in the HIDs presented above corresponds to a) periods with high flux variability (by a factor close to 3) at almost 
constant hardness and b) periods where the flux intensity and hardness are strongly anti-correlated. The second result is that the low energy
 emission evolves drastically in terms of luminosity while at higher energy, the emission follows  a less extreme pattern.
From HID and spectral studies, a reasonable scenario is based on a low energy component attributed to a Comptonizing population and  second component,
 described by a cut-off powerlaw shape, extending  up to $\sim$ 600 keV, beyond the Comptonization cut-off. 
The two components coexist and evolve together but not in a strictly correlated way.  On a few hour timescales, it appears
that the first flare occuring at the beginning of revolution 1557 
is dominated by the hard component, while the major flare is driven by the low energy component. As an important radio flare
was reported around 2015, June 26,17:00 (MJD 57199.71, Tsubono et al. 2015), i.e. between INTEGRAL revolution 1556 and 1557,
it is possible that we observe a transition from a jet (outflow) dominated to an accretion (inflow) dominated regime, even if both components 
are present all along the active period. It differs from the behavior observed during the Soft/Hard state transition, where energy 
release seems to be transferred from disk to corona (or conversely). Here, the jet and the accretion flow  are to be considered together, 
as possible contributors to the high energy emission, as concluded from previous works on Cyg X-1 for instance (Laurent et al. 2011, Jourdain et al. 2012).
Still we cannot exclude that the hard tail above the (low energy) Compton emission is due to a second hotter Compton population (temperature around 100 keV)  or to some gradient in space and/or time of the electron temperature inside the plasma region (e.g. Skibo \& Dermer, 1995; Malzac \& Jourdain 2000), which may be further complicated by a super-Eddington regime.\\
Whatever the nature of these emissions, the huge variability (of both components), observed on a 5 minutes timescale, contains 
 information related to the underlying mechanisms and to the 
emitting region characteristics  (size, electron population distribution or magnetic field configuration,...). 
In addition to another important outburst in 1989, V404 Cyg is known to vary drastically, even in  quiescence. 
It is worth noting, for instance, that a similar variability is observed in radio during its quiescent state (slope change on a 10 minute timescale) (Rana et al. 2016).

In the  scenario proposed in this paper, the Comptonizing population parameters evolve along the flare within 5 minutes.
At the top of the major flare, the  temperature remains constant during a large part of the flare (constant hardness) and increases regularly 
after the maximum, accompagnying the low energy flux decrease (kT from $\sim$  20 keV to 30 keV in the post-flare phase). Such a behavior can be explained by  
a decreasing soft photon flux entering the corona, due to a lower accretion rate. 

\begin{figure}
 \includegraphics[width=8cm,height=8cm,angle=90]{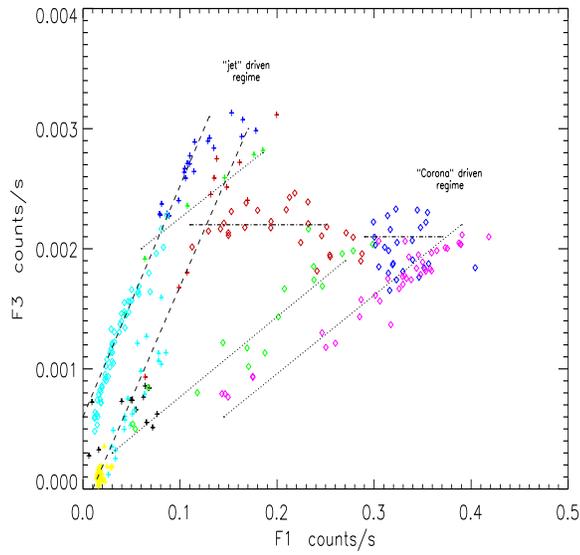}\\
\caption{100-300 keV flux versus 20-50 keV flux;  Lines correspond to the global trends discussed 
in the text.}
\label{fig:F3F1}
\end{figure}

The second component cannot be strongly constrained on the short timescale used here. We fixed the photon index to 1.6 and, for most of the spectra, a common cutoff energy of 150 keV
when not constrained by the data. The important conclusion comes from the normalization evolution and its potential relation with the low energy component.
To investigate this point, we  plot the relation between F1 and F3, as tracers respectively of the low and high energy component (Fig. \ref{fig:F3F1}).
Note that "+" symbols corresponds to first flare and diamonds to second one.
Three kinds of correlations appear,  vizualized in the plot by three families of straight lines.\\
1) R1 corresponds to F3 = $\alpha$ $\times$ F1 + $\beta_{i}$, with i=1,2 \\
2) R2 corresponds to F3 = $\alpha$'$\times$ F1 + $\beta'_{i}$, with i=1,2,3\\
3) R3 corresponds to F3 = Constant\\

With this very simplified scheme, we have a global view of the source behavior all along the revolution. 
Basically, the source follows the R1-type relation during the first flare, (on the left of the figure), 
and the R2-type relation during the second flare. In both cases, the points are split into two groups, 
following parallel lines with different offset terms ($\beta$). $\beta$ decreases between the two 'bumps' of the first flare
and $\beta'$ decreases between the rise phase and the second bump for the second flare. Still, the first bump marks the transition between them.\\ 
The decay phase appears clearly, with this representation, as a decrease of the first spectral component ($\sim$ F1 flux)  with an almost constant
second spectral component ($\sim$ F3 flux; R3-type relation). More striking, this evolution indicates a transition from a R2-type to a R1-type relation.\\
Then, during the final phase, the emission  extends the R1-type track drawn by the first bump of the first flare, potentially
 indicating the come back to a same source configuration.\\
In addition, we can note that the rise phase of the first bump (yellow '+' in the left) seems to follow
a R2-type relation (but the number of points is limited).

Even if the source behavior can not be described by a unique relation (which would have been surprising, given the intensity of the event), the above 
diagrams (HID or flux-flux) demonstrate that the emission process is not so chaotic and  is driven by a competition between two mechanisms emitting 
 high energy photons. The two axes observed in Fig.\ref{fig:F3F1} identify two regimes : one with  the low energy component dominating the source evolution; the second where the high energy component contributes significantly to the total emission.  We observe also a transition between those different configurations after the maximum of energy release.
To go further, we point out that many radio observations have reported the formation of a (variable) jet during the ten days or so
of the V404 Cyg outburst. 
If we identify the low energy component with the corona emission and the flux above 100 keV with a jet contribution, the two main relations, R1 and R2,
illustrate  how both components compete for producing the hard X-ray emission. 

A last point to address is the unusally hot seed photon temperature (5-7 keV) required to get acceptable $\chi2$ values.
Scenarii with very thick (of the order of $10^{24}~cm^{-2}$) absorbing medium  and/or high reflection factor (a few) 
could also reproduce the spectral shape. 
In this case,  
the observed soft X-ray emission (below 10 keV) could escape the absorption thanks to a partial covering 
of the emitting region or to a different production site.
Due to the super Eddington regime reached by the source, a dense outflow is expected to form, supporting  
the presence of very dense absorbing clouds, randomly occulting part of the source emission. However, 
 such a scenario is able to explain the spectral but not the intensity varibility above 20 keV.
For instance,  episodes with flux evolution at constant  hardness cannot be due to
a change in the column density. Conversely, if present, the absorbing material configuration
should be stable in time or less than $10^{24}~cm^{-2}$.\\
We have tested such a model to check its influence on the source parameters. 
Tab.\ref{tab:fitsNH} gives the results of the fits of the same spectra as Tab.\ref{tab:fits}, with the same two component model absorbed by a very thick material. The photon seed temperature has been fixed to 1 keV, the cutoff power-law index to 1.6,  and cutoff energy to 150 keV when not constrained.
The column density is a key parameter for the spectral shape at low energy and the underlying Compton component  
will be affected. Degeneracy between both curvatures (absorption and Compton) may limit our conclusion. Nevertheless, we obtain a good description of our data with this kind of scenario, and comparable parameters.
In particular, the spectral evolution
 during the decay  is still driven by an increase of the Compton parameter and compatible with a constant absorption value.\\
Note that if an absorbing material surrounds the source, it should also produce an important reflected
 component. Including all the ingredients to describe the possible geometries (Nh, R, covering
 fraction, ...) will increase the number of  degrees of freedom  and prevent any firm conclusion. 
Though, even if the values of the parameters will depend on the model hypotheses,
the global trends brought to light in this analysis will remain.

\section{summary and conclusion}
 
We present  INTEGRAL/SPI  observations of the major (and final) flaring episode exhibited by the X-ray transient V404 Cygni during its
 very bright 2015 outburst.
The spectral variability has been studied on a timescale of 5 minutes in the hard X-ray domain (above 20 keV). 
From the HID, we bring to light two main variability schemes:
On one side, we observe huge amplitude variations (by a factor of 4-5)  with the source emission keeping an almost constant spectral shape and low hardness.
On another side, the source luminosity is anti-correlated to the hardness, in particular during the decay phase observed at low energy after the brightest peak of luminosity.

The light curve and hardness evolution patterns can be explained by the presence of two components in the spectral emission, which are not strictly correlated
but reveal an organized underlying picture.
A spectral fit with a Comptonization component plus a cutoff power-law is used to extract some physical information on the emission characteristics.
In addition, looking at the relation between low energy  (20-50 keV) and high energy fluxes (100-300 keV), we find two kinds of relations, which  
point out two different regimes in the source energy release, probably linked to the nature of the underlying emission mechanisms for the high energy photons.
Considering the two main potential power sources, accretion and magnetic field, 
which act on one  another, it is  important  to identify the respective role of the observed spectral components, as well as their interconnection.\\
Moreover, each of these components evolves intrinsically along the flare. In particular, the Compton component
itself behaves in two specific manners, presenting both
large amplitude variations at constant spectral shape (parameters values) and hardness-flux anticorrelation episodes, where the temperature and optical depth 
of the Comptonizing population increase when the flux decreases.\\
The V404 Cyg 2015 outburst is a unique opportunity to explore in depth the spectral evolution of an X-ray binary system.
It  provides us the best chance to get a clearer idea
of the interplay between the two hard X-ray components already observed in a couple of bright/hard objects (Cyg X-1, 
GX 339-4 being the most famous), even if this spectacular event still requires additional investigation. 
The large amount of data available on this event will be used in futher works to go in more details in the source comportment.

The complexity of the source behavior remains an important challenge, but the above analyses have revealed an 
organized  picture of two spectral components that are
significant steps on the road of a better understanding
of the nature of the high energy emission of compact objects.
\\
\\
\textit{The V404 Cyg results from INTEGRAL/SPI are available at}
https://sigma-2.cesr.fr/integral/spidai.
 \textit{It contains the scientific results from Roques et al. (2015) on INTEGRAL revolution 1554 and this work, in particular the light curve fluxes (Fig. 1) and hardness ratios (Fig. 2 and 3).}
 
\section*{Acknowledgments}  The \textit{INTEGRAL} SPI project has been completed 
under the responsibility and leadership of CNES. We are grateful to ASI, CEA, CNES, DLR, ESA, INTA, NASA and OSTC for support.


\newpage
\begin{deluxetable}{lccccccccccc}
\tablewidth{0pt}
\tablecaption{Best fit parameters for a sample of V404 Cyg spectra .}
\tabletypesize{\scriptsize}
\tablehead{ 
\colhead{Spectrum}
&\colhead{$T_{start}-T0$}
&\colhead{$kT_0$} 
&\colhead{kT} 
&\colhead{$\tau $}
 &\colhead{$E_{cut}$}
&\colhead{$\chi_{red}^{2} $}\\
&\colhead{H}
  &\colhead{keV}
&\colhead{keV}
  &&\colhead{keV}
&\colhead{(DoF)} \\
%
 }
\startdata 
2nd bump of 1st flare & 220.965&   7.2 $\pm$ 0.1 & 25.2 $\pm$ 2 &  1.45 $\pm$ 0.1     & 150 (fix)       &  1.02 (33)    \\
Rise Phase of 2nd flare& 227.037 & 5.9  $\pm$ 0.1 & 20.1 $\pm$ 1 &  1.26 $\pm$ 0.08     & 150 (fix)       &  1.03 (33)    \\
2nd bump of 2nd flare  & 230.800 & 5.2 $\pm$ 0.1& 21.0$\pm$ 1 & 1.0  $\pm$ 0.1   & 229 $\pm$  40        &2.1 (32)    \\
=  decay == &&&\\ 
Decay 1 & 233.644 & 6.35 $\pm$ 0.1 & 20.6$\pm$ 2 &  1.23 $\pm$ 0.15     & 150 (fix)        &  2.0 (33)   \\
Decay 2 & 235.638& " & 21.0 $\pm$ 1 &  1.71 $\pm$ 0.2      & 150 (fix)      &  1.1 (33)   \\
Decay 3 & 236.511 & " & 22.4 $\pm$ 1.5  &  2.1 $\pm$ 0.2   & 150 (fix)       &  0.7  (33)   \\
 =  final part  ==&&&\\ 
Mean Spectrum  (4) & 236.595 & 7.0 (fix)  &   29.8 $\pm$ 1 &   2.2 $\pm$  0.1       &   283  $\pm$  30       &  2.2 (33)   \\
Indiv. Peak (5)& 238.248 &  "   & 26.5$ \pm$ 2 & 2.2  $\pm$ 0.3        & "         &  1.2 (34)   \\
\enddata
\tablecomments{Results of spectral fits for spectra displayed in Fig.\ref{fig:SoftHard} (first three) 
and Fig.\ref{fig:decay} (decay and final part).
The model consists in a Comptonization component and a cutoff power-law (compTT+cutoffPL in Xspec). The latter has a fixed slope (1.6). Its cutoff energy has also been fixed
when not constrained.  0.5\% of systematic have been added to the data for all spectra.
}
\label{tab:fits}
\end{deluxetable}

\begin{deluxetable}{lccccccccccc}
\tablewidth{0pt}
\tablecaption{Best fit parameters for a sample of V404 Cyg spectra .}
\tabletypesize{\scriptsize}
\tablehead{ 
\colhead{Spectrum}
&\colhead{$Nh$} 
&\colhead{kT} 
&\colhead{$\tau $}
 &\colhead{$E_{cut}$}
&\colhead{$\chi_{red}^{2} $}\\
  &\colhead{$10^{22}~cm^{-2}$}
&\colhead{keV}
  &&\colhead{keV}
&\colhead{(DoF)} \\
%
 }
\startdata 
2nd bump of 1st flare  &  710 $\pm$ 60 & 27.2 $\pm$ 3 &  1.2 $\pm$ 0.1     & 150 (fix)       &  0.98(33)    \\
Rise Phase of 2nd flare  & 667 $\pm$ 60 & 23.6 $\pm$ 2 &  0.83 $\pm$ 0.1     & 150 (fix)       &  1.1 (33)    \\
2nd bump of 2nd flare  & 557 $\pm$ 60 & 26.4$\pm$ 2 & 0.6  $\pm$ 0.1    & 265 $\pm$  40        &2.1 (32)    \\
=  decay == &&&\\ 
Decay 1  & 581 $\pm$ 100 & 19.5$\pm$ 2 &  1.2 $\pm$ 0.2      & 150 (fix)        &  1.98 (33)   \\
Decay 2 & 480$\pm$ 100 & 21.8 $\pm$ 2 &  1.45 $ \pm$ 0.2      & 150 (fix)      &  1.05 (33)   \\
Decay 3  & 545 $\pm$ 100 & 27.6  $\pm$ 3.5  & 1.4 $\pm$ 0.2   & 150 (fix)       &  0.6  (33)   \\
 =  final part  ==&&&\\ 
Mean Spectrum  (4) & 340 $\pm$ 60 &   29.0 $\pm$ 3 &   2.24 $\pm$  0.5       &   206  $\pm$  40       &  2.0 (33)   \\
Indiv. Peak (5) &  "  & 25.2 $ \pm$ 1.5 & 2.26  $\pm$ 0.2        & "         &  1.2 (34)   \\ 
\enddata
\tablecomments{Same as Table \ref{tab:fits} for a model consisting of absorbed Comptonization component and  cutoff power-law (wabs*(compTT+cutoffPL) in Xspec). Seed photon temperature for Comptonization is fixed to 1 keV. The cutoff power-law has a 1.6 fixed slope as well as cutoff energy when not constrained. 0.5\% of systematic have been added to the data.
}
\label{tab:fitsNH}
\end{deluxetable} 


\end{document}